\documentstyle [12pt] {article}  \headheight  =  4mm
= 230mm \textwidth = 165mm \normalbaselineskip = 12pt \normalbaselines
\begin  {document} \begin {center} {\bf ON THE POSSIBILITY OF LOCAL SR
CONSTRUCTION} \\ [4mm] {\bf G.A. Kotel'nikov} \\ {\it Russian Research
Centre   Kurchatov   Institute,  Moscow  123182,  Russia}  \\  E-mail:
kga@electronics.kiae.ru \end {center} \begin {abstract} The  violation
of the invariance of the speed of light in Special Relativity has been
made.  The version of the theory has been  constructed  in  which  the
possibility of the superluminal motions are permitted. \end {abstract}
\section {Introduction} \label {1} In view of  mathematical  elegance,
laconicalness  and  predictive  power  Special  Relativity (SR) is the
fundamental theory of modern physics.  Owing to this the  mathematical
postulates  of  the  theory,  possibility  of  their  modification and
generalization as well  as  of  experimental  test  attract  attention
constantly.  As  examples  one  can  be presented the well known Pauli
monograph \cite {Pau47},  containing the elements of the  Abraham  and
Ritz  theories;  academician  Logunov's lectures on the foundations of
Relativity Theory with the formulation of SR in the affine space \cite
{Log82};  Fushchich's  publication  on  the non-linear electrodynamics
equations  with  the  non-invariant  speed  of  light  \cite  {Fus92};
Glashow's  work  on  the experimental consequences of the violation of
the Lorentz-invariance in astrophysics \cite{Gla99}.

To the  present  time  SR  is one of the most experimentally-justified
theories (for example,  Pauli and Landsberg monographies  \cite{Pau47,
Lan76};   Strakhovsky   and   Uspensky  \cite{Str65},  Basov  and  his
co-authors \cite{Bas61},  M\o ller \cite{Mol62}  and  Molchanov  \cite
{Mol64}  reviews;  the  original  publications of \cite{New78,  Com79,
Coo79}).  Here one can mention the experiments  on  detection  of  the
ether  wind  in  the  experiments  of  the Michelson type \cite{Pau47,
Lan76};  determination of the angular light  aberrations  \cite{Pau47,
Lan76};   transversal   Doppler   effect   measurement   \cite{New78};
experiments on the proof of independence of the speed  of  light  from
the velocity of the source of light \cite {Mol64,  New78}; experiments
on determination of the relativistic mass dependence of  the  velocity
of  a  particle motion \cite {New78};  the relativistic retardation of
time \cite {New78};  the g-2 experiments \cite{Com79,  Coo79,  New78}.
The  results  of  these  experiments indicate the absence of the ether
wind to closer and closer limits of accuracy, and argue for SR.

This raises  the  natural  question,  whether  do  exist  at  all  any
experiments different  from  SR  predictions,  even  though  they  are
ambiguously  interpreted.  It  appears  that there are a number of the
publications on this theme.  Let  us  consider  those  concerning  the
second  postulate  -  the  postulate  of the constancy of the speed of
light.

Giannetto, Maccarrone,  Mignani and Recami  \cite  {Gia86}  have  been
considered  the possibility of the negative sign interpretation of the
square of the neutrino 4 - momentum $ P ^ 2 = E ^ 2 - {\bf p} ^ 2 c  ^
2  =  {M  _  0}  ^  2  c  ^ 4 = (-0,166\pm 0,091) $ $ MeV ^ 2 $ in the
experiments on $\pi$ - decay $ \pi ^ + \to\mu ^ + + \nu $ as the  fact
of  observation of a superluminal particle with imaging mass $ M _ 0 =
im $ (tachyon).  Khalfin \cite {Kha96} has established  that  negative
sign  of  the  square  of  the  neutrino  4-momentum  may  be  due  to
incorrectness of the observational data processing near upper bound of
$  \beta $ - spectrum (in our own case near a upper bound of $ \mu $ -
spectrum).  Thus,  the possibility of  the  interpretation  of  $\pi$-
neutrino as a particle of the tachyon nature is eliminated practically
in the light of the contemporary explanation for the negative sign  of
the 4-momentum square.

Mamaev \cite  {Mam93} has analyzed the time-flight spectra of $\pi^-$,
$\mu^-$,  $e^-$ particles from Joint Institute  for  Nuclear  Research
(Dubna)  and concluded that the data from the article \cite{Bun78} may
be interpreted as the result of  superluminal  motion  of  mesons  and
electrons.  However,  taking  into  account the presence in the signal
processing electronic circuit of a  threshold  device  (discriminator)
with  2543  channels of the analyzer,  it is possible to conclude that
the velocities of these particles were $0,92c $, $ 0,94c $ and $ 0,96c
$  respectively,  where  $c$ is the speed of light.  The phenomenon of
superluminal motion disappears, and mutual arrangement of the spectral
lines  from \cite{Bun78} may be explained in the framework of SR \cite
{Kot68}.

Nevertheless numerous examples are known in which the  elimination  of
superluminal motion turns out to be more difficult and less convincing
than in the considered cases.  These are observations of  superluminal
motion  of  particles  in  broad  atmospheric  showers and the acts of
antiproton birth,  as well as on  expansion  of  the  shells  of  some
extragalactic radiosources, for example \cite{Cla74, Coo79, Mat83}.

Clay and  Crouch  has  observed \cite {Cla74} impulses,  preceding the
signal induced by a broad atmospheric  shower.  Let  us  suppose  that
particles from the shower had the velocity equal to the speed of light
(that is natural).  Then it is not  clear,  what  has  preceded  these
particles.  "We  conclude  that  we  have  observed  non-random events
preceding the arrival of an extensive  air  shower.  Being  unable  to
explain this result in a more conventional manner,  we suggest that is
the result of a particle traveling with an apparent  velocity  greater
than  of light " \cite{Cla74}.  Further the authors \cite {Cla74} have
assumed that the  impulses  were  stipulated  by  the  particles  with
imaginary  masses (tachyons) traveling at the velocities exceeding the
speed of light.

Cooper \cite {Coo79} has concluded that the time-flight experiments on
observation   of  antiprotons  admit  the  existence  of  superluminal
particles (antimesons)  connected  with  antiprotons.  The  calculated
probability  of  the  velocity  of  antimesons  exceeding the speed of
light,  is equal 0.9972.  The evaluation turn out to  be  tolerant  to
various experimental errors.  The author writes:  " A reexamination of
the  Nobel-prize-winning  experiment  in  which  the  antiproton   was
discovered  reveals  that  associated  antimesons  might  be traveling
faster than light " \cite{Coo79}.

The numerous publications are known on the observation of superluminal
expansion  of  extragalactic  radiosources  (for example \cite {Mat83,
Gin87,  Kot68}).  It is an interesting phenomenon, and it is difficult
to  be  explained in terms of modern astrophysics.  The observation of
the  superluminal  expansion   became   possible   after   the   radio
interferometers  VLBI  (Very  Long  Baseline  Interferometry)  for the
centimetre spectral range were  created.  These  possess  a  superlong
trans-continental   base   $L$   (thousands   and  tens  of  thousands
kilometers).   The   angular   resolution   of   such   telescopes   $
\delta\sim\lambda/L  $  is  proportional  to  the  ratio  of a working
wavelength $ \lambda $ to the value of the base $L$. It is much higher
than  the  one  of  the  best optical devices.  In the optical range $
L/\lambda $ is equal $ \sim 6\cdot10 ^ 7 $,  while in the radio  range
it  is equal $ \sim 18\cdot 10 ^ 8 $.  The radio interferometers allow
one to study such thin structure of space objects ($ \sim 7\cdot 10  ^
{-4}  $  angle  seconds) as was inaccessible to be observed by optical
means.  The studies have shown that many extragalactic  radio  sources
have   a  complicated,  bi-component  structure.  Among  of  them  the
substructure of  six  radio  sources  run  away  from  each  other  at
calculated  velocities  that  are  some  times  more than the speed of
light.  It is the radio galaxy 3C120 (z = 0.033),  quasars 3C273 (z  =
0.158),  3C279 (z = 0.538),  3C345 (z = 0.595),  3C179 (z = 0.846) and
NRAO140 (z = 1.258) \cite {Sci81}.  (Here  $z$  is  the  parameter  of
redshift).  The transversal velocities calculated within the framework
of the cosmological Friedmann model of the motion  of  the  components
are equal $ V _ {\perp} \sim (2-20) c $. It has been proposed over ten
versions for interpretations of this phenomenon.  It may be associated
with  more  complicated  multicomponent structure of the quasars;  the
random superposition of radio  spots  on  the  quasars;  influence  of
intergalactic gravitational lens duplicating a visible image;  Doppler
effect;  increase  of  Hubble's  constant  that  is   accompanied   by
decreasing the distances to the quasars, which results in disappearing
the superluminal expansion.  Also,  it may be due to the influence  of
interstellar   magnetic   fields;   existence   of   tachyon   matter;
introduction of 5-space with an additional fifth  coordinate  such  as
the  speed of light running the values from $0$ to $\infty$;  model of
the light echo;  optical illusion not contradicting to SR \cite{Mat83,
Gin87, Kot68}. It is evident that the conventional explanation for the
superluminal expansion is not offered yet, and various hypothesizes on
the nature of this phenomenon may be discussed.

Loiseau \cite  {Loi68}  has  paid  attention  to the little difference
between  the  galaxy  NGC  5668  redshift  $  z  '  $,   measured   by
radioastronomical   method  at  the  frequency  corresponding  to  the
wavelength 21 cm,  and the redshift $z$, measured in the optical range
for this galaxy.  This result, {\it if it really is outside the limits
of measurement errors}, cannot be explained in the framework of SR, as
$ z ' = z $ should be with $ c ' = c $. To explain this result, author
\cite{Loi68} introduced 3-dimensional  non-Euclidean  space,  inserted
into  4-dimensional  Riemannian  space with some common time.  In this
case it may be obtained that the galaxy speed of light  $c'$  and  the
speed  of  light $c$ on the Earth are connected by the ratio $ c ' = c
(1 + z) / (1 + z ') $, where $ z $ is the redshift on a wave length in
the optical range,  and $ z ' $ is the redshift on a wave frequency in
the radio range.  In accordance with the observed data on  the  galaxy
NGC  5668 $z$ is equal to 0.00580 in the optical range;  $z'$ is equal
to 0.00526 in the radio range on the frequency  corresponding  to  the
wavelength  21 cm.  It follows from here that $ c '/c = (1 + z) / (1 +
z') = 1.00580/1.00526 = 1.0005372 $, and $ c ' = c + 182,04 $ km / sec
>  $  c  $  \cite {Loi68}.  The estimation has shown that the speed of
light from quasar QSO PKS 2134 with the optical redshift $ z = 1.936 $
is equal to $ c ' = 440.000 $ km / sec \cite {Loi68}. The relationship
between $ c ' $,  $ c $ and the quasar velocity $ v $ relative to  the
Earth  is  described  by  the  formula $c'=c\sqrt{1 + v^2/c^2}$ in the
approximation  of  a  weak  gravitational   field.   The   statistical
significance of the hypothesis on the difference between the redshifts
in the radio and optical ranges is naturally the deciding  factor  for
the Loiseau work. \bigskip

Thus, unambiguously  interpreted  experimental  data  distinct from SR
predictions are apparently absent now. But there are vague indications
that  it  is not improbable that they exist in particle physics and in
astrophysics.  Let us consider the hypothesis on the existence of  the
superluminal  motion  in  terms  of the violation of invariance of the
speed of light in the expression for the second degree  of  4-interval
at the infinitesimal level.

\begin {sloppypar}   \noindent   \section   {Space   -   Time  Metric,
Differentials  Coordinates  Transformation   Law}   \end   {sloppypar}
\noindent  Let us start from the condition for the invariance of the 4
- interval differential in Minkowski space  with  the  metric:  \begin
{equation} \label {g49} \begin {array} {c} ds ^ 2 = - ({dx _ 1} ') ^ 2
- ({dx _ 2} ') ^ 2 - ({dx _ 3} ') ^ 2 - ({dx _ 4} ') ^ 2 = \\ - (dx  _
1)  ^  2  -  (dx  _  2) ^ 2 - (dx _ 3) ^ 2 - (dx _ 4) ^ 2 - inv.  \end
{array} \end {equation} Here $ dx _ {1,2,3} = (dx, dy, dz) $, \ $ dx _
4  =  icdt  $,  it  is not necessary for the speed of light $c'$ to be
equal $ c $. Corresponding infinitesimal space - time transformations,
saving the invariance of the form (\ref {g49}),  obviously contain the
group locally isomorphic to the Lorentz group  \cite  {Lan60}:  \begin
{equation}  \label {f2} dx _ a ' = dx _ a,  \ dx _ a ' = L _ {ab} dx _
b,  \ a, b = 1,2,3,4, \end {equation} where $ L _ {ab} $ is the matrix
of  the  six-dimensional  Lorentz group $L_6$ \cite {Lan60} with local
kinematics parameter $  \beta  $.  The  one-dimensional  infinitesimal
transformations corresponding to the given matrix, take the well known
form:  \begin {equation} \label {f3} dx _ 1 ' = \frac {dx _ 1 + i\beta
dx _ 4} {\sqrt {1-\beta ^ 2}};  \ dx _ 4 ' = \frac {dx _ 4-i\beta dx _
1} {\sqrt {1-\beta ^ 2}};  \ dx _ 2 ' = dx _ 2;  \ dx _ 3 ' = dx  _  3
\end   {equation}   \begin   {sloppypar}   \noindent   The  reciprocal
transformations may be obtained by the prime  permutation.  The  group
parameters  are  connected  by  the  ratio  $ \beta ' = -\beta $ \cite
{Lan60}.  But contrary to the  global  Lorentz  transformations  \cite
{Lan60},  here  the  parameters  $  \beta $ and $ \beta ' $ can depend
explicitly or implicitly on a space -  time  point  $  \beta  =  \beta
(f({\bf x}, t)), \ \beta ' = \beta ' (f ' ({\bf x} ', t ')) $. This is
the important circumstance which  will  allow  one  to  construct  the
theoretical  model  in  which  the existence of superluminal motion is
possible.  The integral space - time transformations induced by  (\ref
{f3})  are:  \end  {sloppypar}  \begin  {equation}  \label {f4} \begin
{array} {c} \vspace {1mm} \displaystyle x _ 1 ' = \int\frac  {dx  _  1
+i\beta  dx_4} {\sqrt {1-\beta ^ 2}} + d _ 1;  \displaystyle x _ 4 ' =
\int\frac {dx _ 4 -i\beta dx_1} {\sqrt {1-\beta ^ 2}} + d _ 4;  \\ x _
2  '  =  x _ 2 + d _ 2;  \ x _ 3 ' = x _ 3 + d _ 3,  \end {array} \end
{equation} where $ d _ 1 - d _ 4 $ are the translation parameters; the
reciprocal transformations may be obtained by the prime permutation; $
d _ a ' = -d _ a $,  $ a = 1,2,3,4 $.  The transformations (\ref {f4})
go  into the Poincar\'e ones if $ c $ = cost,  $ c ' = c $ be put into
them and our consideration be restricted to inertial motions ($  \beta
$  =  const).  In  this  case on integration they go into the standard
transformations from Poincar\'e group (inhomogeneous  Lorentz  group).
Thus,  Lorentz  transformations  are  contained here as the particular
case. The group properties of the integral transformations (\ref {f4})
are   realized  due  to  the  group  properties  of  the  differential
transformations (\ref {f3})  and  due  to  the  relativistic  velocity
addition theorem $ \beta " = (\beta + \beta ') / (1 + \beta\beta ') $.
\bigskip

\section {Integral  of  Operation,  Energy,   Momentum}   \label   {4}
\noindent  Let  us  turn  to  the  integral  of  operation in SR \cite
{Lan60}.  It is not invariant with respect to the transformations with
broken invariance of the speed of light.  However this property may be
corrected if we start from the invariant integral of  operation  \cite
{Kot68}:  \begin  {equation} \label {2.26b} \begin {array} {c} \vspace
{1mm} \displaystyle S ^ * = cS = - mc ^ 2\int ds + e\int A _ a dx _  a
+  \frac  {i}  {16\pi}  \int {F _ {ab}} ^ 2 d ^ 4 x = \\ \vspace {1mm}
\displaystyle -mc ^ 2\int ds-i\int A _ a j _ a d ^ 4  x  +  \frac  {i}
{16\pi}  \int  {F  _  {ab}} ^ 2 d ^ 4 x = \\ \displaystyle \int (-mc ^
2\sqrt {1-\beta ^ 2} + e {\bf A} \cdot\beta\displaystyle e\phi)  (cdt)
+ \frac {1} {8\pi} \int (E ^ 2-H ^ 2) d ^ 3x (cdt).  \end {array} \end
{equation} \noindent Here $ S ^ * $ is the new integral of  operation,
which  we  name  the  generalized  one;  $  mc  ^ 2 $ is the invariant
combination corresponding to the rest energy of a particle ($ m  $  is
the  rest-mass,  $ c $ is the speed of light);  $ e $ is the invariant
electrical charge of a particle;  $ A _ a = (A _ 1,  A _ 2, A _ 3, A _
4) = ({\bf A}, i\phi) $ is the 4-potential \cite {Lan60}; $ j _ a = (j
_ 1, j _ 2, j _ 3, j _ 4) = (\rho {\bf v} /c, i\rho) $ is the 4-vector
of  current  density \cite {Pau47} instead of $ j _ a = (\rho {\bf v},
ic\rho) $ \cite {Lan60},  $ \rho $ is the charge density,  $ {\bf v} $
is the velocity of a charge;  $ F _ {ab} = \partial A _ b/\partial x _
a - \partial A _ a/\partial x _ b $ is the tensor  of  electromagnetic
field;  $  {\bf E} = - (1/c) \partial {\bf A} /\partial t-\nabla\phi $
is the electrical field;  $ {\bf H} = \nabla {\rm X} {\bf A} $ is  the
magnetic field; $ {F _ {ab}} ^ 2 = 2 (H ^ 2-E ^ 2) $; $ d ^ 4 x = dx _
1 dx _ 2 dx _ 3 dx _ 4 $ is the  element  of  the  invariant  4-volume
\cite {Lan60}.

The transformational  rest-mass  properties  is  changed as the result
from the introduction of the generalized integral  (\ref{2.26b}).  The
mass is not any more scalar.  The mass is transformed according to the
law $m'=(c^2/{c'}^2)m=\gamma^{-2}m$.  The rest  energy  $mc^2$  has  a
scalar  property.  The  transformational  property  of  Plank constant
$\hbar$ is changed as well.  The invariant is not the constant $ \hbar
$,  but the product $ \hbar c $. Due to the electrical charge property
of invariance $ e $,  the thin structure constant remains invariant  $
\alpha  =  e  ^ 2/\hbar c $ - inv.  \begin {sloppypar} The generalized
Lagrangian,  energy and 4 - momentum of a particle correspond  to  the
generalized  integral  of  operation.  We  will  label the generalized
values with the symbol *.  We have: \end {sloppypar} \begin {equation}
\label  {2.27}  L  ^  *  = cL = -mc ^ 2\sqrt {1-\beta ^ 2} + e {\bf A}
\cdot\beta-e\phi;  \end {equation} \begin {equation} \label {2.27a}  {
\bf  P}  ^ * = \frac {\partial L ^ *} {\partial\beta} = \frac {cm {\bf
v}} {\sqrt {1-\beta ^ 2}} + e {\bf A} = c {\bf p} + e  {\bf  A};  \end
{equation}  \begin  {equation}  \label  {2.28b}  E  ^  *  =  {\bf P} ^
*\cdot\beta-cL = \frac {mc ^ 2} {\sqrt {1-\beta ^ 2}}  +  e\phi  =  E.
\end {equation} It follows from here that the motion integrals are the
energy $ E $ and the product of the speed of  light  by  the  momentum
from SR:  $ c {\bf P} = c {\bf p} + e {\bf A} $. The parameter $ \beta
$ has meaning as generalized velocity. The differential $ dx ^ 0 = cdt
$ plays a role of the time differential. It is essential that owing to
the differentiation with respect to  the  parameter  $  \beta  $,  the
results   obtained   do  not  depend  on  the  particular  assumptions
concerning the properties of the speed of light,  as the value $  c  $
enters into the parameter $ \beta = v/c $. \begin {sloppypar} Owing to
the well known property of 4 - speed $ U ^ 2  =  -1  $,  we  have  the
following  expression for the generalized 4 - momentum $ {p _ a} ^ * =
mc ^ 2 u _ a $ of a free particle:  \end {sloppypar} \begin {equation}
\label  {2.30}  {  {p _ a} ^ *} ^ 2 = c ^ 2p ^ 2-E ^ 2 = -m ^ 2c ^ 4 -
inv.  \end {equation} As in \cite {Lan60},  in case of a  particle  in
electromagnetic  field we find:  \begin {equation} \label {2.31} { P _
a} ^ * = mc ^ 2 u _ a + eA _ a; \end {equation} \begin {equation} ( {P
_  a} ^ *-eA _ a) ^ 2 = (cP _ a-eA _ a) ^ 2 = -m ^ 2c ^ 4 - inv.  \end
{equation}

\begin {sloppypar} \noindent \section {Equations of Motion for Charged
Particle}  \end  {sloppypar}  \label  {5} \begin {sloppypar} \noindent
Keeping in the mind expression (\ref  {2.27}),  we  shall  start  from
Lagrange equations $ d (\partial L ^ */\partial\beta) /dx ^ 0-\partial
L ^ */\partial {\bf x} = 0 $ taking into account the vector equality $
\nabla ({\bf a} \cdot {\bf b}) = ({\bf a} \cdot\nabla) {\bf b} + ({\bf
b} \cdot\nabla) {\bf a} + {\bf a} {\rm x} ( \nabla {\rm x} {\bf b})  +
{\bf  b}  {\rm x} (\nabla {\rm x} {\bf a}) $ \cite {Lan60}.  We obtain
the following equations for  the  motion  of  a  charged  particle  in
electromagnetic  field:  \end  {sloppypar}  \begin  {equation}  \label
{2.33} \frac {d {\bf p} ^ *} {dt} = \frac {d (c {\bf p})}  {dt}  =  ce
{\bf E} + e {\bf v} {\rm x}{\bf H};  \end {equation} \begin {equation}
\label {2.34} \frac {dE ^ *} {dt} = \frac {dE} {dt} = e {\bf E}  \cdot
{\bf v}; \end {equation}

\section {Maxwell  Equations}  \label  {6}  \begin  {sloppypar} Let us
start from the  permutational  ratios  of  the  electromagnetic  field
tensor  and the field Lagrange equations $ \partial (\partial {\cal L}
^ */\partial A _ {a,  b})  /\partial  x  _  b-  \partial  {\cal  L}  ^
*/\partial A _ a = 0 $ \cite {Lan60,  Bog73 *} taking into account the
expression $ \partial F _ {ab} ^ 2/\partial A _ {a,  b} = 4F _ {ab}  $
\cite  {Lan60} and the density of the Lagrange function $ {\cal L} ^ *
= c {\cal L} = iA _ a j _ a + (i/16\pi) {F _ {ab}} ^ 2 $.  Here $ A  _
a(x)$ is 4-potential;  $ A _ {a, b} = \partial A _ a/\partial x _ b $;
$ a,  b = 1,2,3,4 $;  $ g _ {ab} = diag (-,  -,  -,  -) $.  In sum  we
have:\end  {sloppypar}  \begin {equation} \label {2.36} \begin {array}
{ll} \vspace {2mm} \displaystyle \nabla {\rm X} {\bf E}  +  \frac  {1}
{c} \frac {\partial {\bf H}} {\partial t} = 0;  & \displaystyle \nabla
\cdot {\bf E} = 4\pi\rho;  \\ \displaystyle \nabla  {\rm  X}  {\bf  H}
-\frac  {1}  {c} \frac {\partial {\bf E}} {\partial t} = \displaystyle
4\pi \frac {{\bf j}} {c};  & \nabla \cdot {\bf H} =  0.  \end  {array}
\end  {equation} \begin {sloppypar} \noindent Out of them the equation
of motion (\ref {2.34}) and the  equations  of  electromagnetic  field
(\ref  {2.36})  coincide  with  the equations which are known from SR.
\end {sloppypar}

Let us  note that according to the given scheme Maxwell equations turn
out to be invariant not only in inertial frames (it is  well   known),
but  also  in non-inertial frames in the flat pseudo - Euclidean space
with the metric $ ds ^ 2 = c ^ 2dt ^ 2-d {\bf x} ^ 2 $.  This property
of Maxwell equations seems to be unusual, but it is known and has been
noted by academician Logunov:  "...  in the  framework  of  SR  it  is
possible  to  describe  a physical phenomena in non-inertial frames as
well.  Fock understood this deeply ...  " \cite {Log82}. The statement
follows  also  from  the  general  covariant  formulation  of  Maxwell
equations \cite {Log82, Lan60, Kot68}.

\begin {sloppypar}  \noindent  \section  {Local  SR}  \end {sloppypar}
\label {8.2} \noindent Up to this point any constraint did not  placed
on  the transformation properties of the speed of light in the theory.
It turns out that  it  is  possible  to  realize  various  theoretical
considerations  by  appropriate  postulation.  In  particular,  if  we
postulate that $ c ' = c $, all the obtained equations will go into SR
equations.  If  we  state  that the speed of light is constant and $ c
'\ne c $, $ c't' = ct $ - inv, the model may be realized which we name
SR {\it with non-invariant speed of light} \cite {Kot68}. It describes
the  same  physical  reality  as  SR  and  also  contains   additional
classification  capabilities  due to the symmetry with respect to more
general group of transformations. Besides Poincar\'e group, this group
includes      the      group     induced     by     the     generators
$X_{-1}=\partial_t-t\partial_t/c,  \  X_0=c\partial_c-t\partial_t,   \
X_{+1}=c^2\partial_c-ct\partial_t$    \cite{Kot68}.   At   last,   the
postulation is possible which permits one to construct  a  version  of
the theory compatible with the principle of relativity and the concept
of superluminal motion. Let us consider this possibility at length.

According to Ritz \cite{Pau47} we assume that the speed  of  light  is
equal  to  $  c _ o = 3\cdot 10 ^ {10} $ cm/sec not in global meaning,
but only relatively to an emitter.  Let us add a new physical  element
to  the  infinitesimal transformations (\ref {f2}) and the model based
on them with equations of motion (\ref {2.33}),  (\ref {2.34}), ( \ref
{2.36}).   We   assume   that   the  state  of  motion  (inertial,  or
non-inertial) does not influences on the proper value of the speed  of
light  $  c  _ 0 $,  Plank constant $ \hbar _ 0 $,  the thin structure
constant $ \alpha $ and other physical proper values, for example, the
proper  length $ l _ 0 $,  proper time $ t _ 0 $,  proper frequency of
oscillations $ \omega _ 0 $,  rest-mass $ m _ 0 $, electrical charge $
e  $.  (A  proper  value  is  the  physical  value  in the frame $K_0$
relatively to which the object is immobile). These remain invariant in
the process of motion: \begin {equation} \label {f} \begin {array} {c}
\vspace {1mm} c _ 0 ' = c _ 0 = c _ o = 3\cdot 10 ^  {10}  cm/sec;  \\
\vspace  {1mm}  \hbar  _  0 ' = \hbar _ 0 = 1,0\cdot 10 ^ {-27} g\cdot
cm^2/sec;  \\ l _ 0 ' = l _ 0;  \ t _ 0 ' = t _ 0;  \ \omega _ 0  '  =
\omega; \ m _ 0 ' = m _ 0; \ e ' = e. \end {array} \end {equation} The
hypothesis on the independence of proper values of physical quantities
from  the state of a physical object motion we agree to name the local
relativity principle.

Let us  assume  further  that  the time intervals measured by means of
differently-placed clocks in any frames $ K,  \ K ', \cdots $ coincide
with  the  local time in a proper frame $ K _ o $ on the trajectory of
the motion of the object:  \begin {equation} \label {t} dt _ o = dt  =
dt'. \end {equation}

We agree to name the theoretical model, realizing the local relativity
principle in the flat space - time with the  metric  (\ref  {g49})  in
combination  with  the  hypothetical  property of time (\ref {t}),  as
Local Special Relativity Theory (LSR) as distinct from  the  classical
SR.  We  find the following expressions for infinitesimal space - time
transformations in this case:  \begin {equation} \label  {g51}  \begin
{array}  {c} \displaystyle dx = \frac {dx _ o-v _ odt _ o} {\sqrt {1 -
{v _ o} ^ 2 / {c _ o} ^ 2}};  \ dy = dy _ o;  \ dz = dz _ o; dt = dt _
o-\frac  {v _ odx _ o} {{c _ o} ^ 2};  \\ \displaystyle c = \frac {c _
o} {\sqrt {1 - {v _ o} ^ 2 / {c _ o} ^ 2}};  \\ \displaystyle dx _ o =
\frac {dx-vdt} {\sqrt {1-v ^ 2/c ^ 2}};  \ dy _ o = dy; \ dz _ o = dz;
\displaystyle dt  _  o  =  \frac  {dt-vdx/c}  {1-v  ^  2/c  ^  2};  \\
\displaystyle  c  _  o = c {\sqrt {1-v ^ 2/c ^ 2}},  \end {array} \end
{equation} where $ v/c = -v _ o/c _ o $;  \ $ (1-v ^ 2/c ^ 2) \cdot (1
+ v ^ 2 / {c _ o} ^ 2) = 1 $.  In this case the following relationship
between the speed of light $ c $ and the velocity of emitter $ v $  is
hold  true  as  the  result  from the transformation properties of the
speed of light in the formulae ( \ref {g51}): \begin {equation} \label
{fc}  \displaystyle  c = c _ o\sqrt {1 + \frac {v ^ 2} {{c _ o} ^ 2}}.
\end {equation} Here $ c _ o $ is the speed of  light  in  the  proper
frame  associated  with  the  emitter.  (In  the  models admitting the
existence of ether $ c _ o $ is  the  speed  of  light  relatively  to
ether).  The  expressions  for  the  speed  of  light  in  the form of
(\ref{g51}) or (\ref {fc})  were  obtained  by  Abraham  (1910)  \cite
{Pau47}  and Rapier (1961) \cite {Rap61} respectively in the framework
of ether models;  next they were obtained by the author of the present
publication  (1968)  \cite  {Kot68} and Loiseau (1968) \cite{Loi68} in
the  framework  of  the  relativity  principle.  These  formulae  were
reproduced  further  by  many authors from various points of view,  in
particular by Marinov  (1975)  \cite  {Mar75},  by  Hsu  (1976)  \cite
{Hsu76},  by  Sj\"odin  (1977)  \cite {Sjo79},  by Mamaev (1985) \cite
{Mam93},  by Nimbuev (1996) \cite  {Nim96},  by  Klimez  (1997)  \cite
{Kli97}, by Russo (1998) \cite {Rus98}. \bigskip

\subsection {Generalized   Momentum   and   Energy}  Let  us  put  the
expression for the speed of light (\ref {fc}) into  the  formulae  for
momentum  (\ref {2.27a}) and energy (\ref {2.28b}) of a free particle.
We have:  \begin {equation} \label {g52} \begin  {array}  {l}  \vspace
{2mm}  \displaystyle  {  \bf  p}  ^ * = c {\bf p} = \frac {cm {\bf v}}
{\sqrt {1-v ^ 2/c ^ 2}} = m _ 0c _ 0 {\bf v};  \\  \displaystyle  E  =
\frac {mc ^ 2} {\sqrt {1-v ^ 2/c ^ 2}} = m _ 0 {c _ 0} ^ 2\sqrt {1 + v
^ 2 / {c _ 0} ^ 2} = m _ 0c _ 0c.  \end {array}  \end  {equation}  The
relation  has  a  view  between the generalized momentum and energy by
this:  \begin {equation} \label {g53} E ^ 2 - {{\bf P} ^ *} ^ 2 = E  ^
2-c  ^  2 {\bf p} ^ 2 = {m _ 0} ^ 2 {c _ 0} ^ 4 (1 + v ^ 2 / {c _ 0} ^
2) - { m _ 0} ^ 2 {c _ 0} ^ 2 {\bf v} ^ 2 = {m _ 0} ^ 2 {c _ 0}  ^  4.
\end {equation}

\subsection {Energy  and  Superluminal  Motion}  Let us begin with the
expression $ v = \sqrt {E ^ 2 - {m _ 0} ^ 2 {c _ 0} ^ 4}  /m  _  0c  _
0\geq  c  _  0 $.  It follows from here that in the framework of LSR a
particle will move with superluminal velocity,  if the particle energy
will  satisfy  the  equality:  \begin {equation} \label {g54} E _ {tr}
\geq\sqrt2 E _ 0 = \sqrt2 m _ 0 {c _ 0}  ^  2.  \end  {equation}  This
energy  is equal $ \sim 720 \ {\rm keV} $ for electron and $ \sim 1330
\ {\rm Mev} $ for proton and neutron.  We may conclude from here  that
neutron   physics  of  nuclear  reactors  may  be  formulated  in  the
non-relativistic approximation in LSR (as in SR).  The electrons  with
the  energy  $  {\it  E} > 720 \ keV $ (for example,  from radioactive
decay) should be superluminal particles in LSR.  Particle  physics  on
modern  accelerators  such as Serpukhov one with the energy of protons
66 GeV (1 Gev = 1000 MeV) should be physics of superluminal motion  in
the framework of LSR, if it would be realized in reality.

\subsection {Equations  of  Motion  for Charged Particle in LSR} After
putting the expression for the speed of light  (\ref  {fc})  into  the
equations  of  motion (\ref {2.33}),  we obtain \cite {Kot68}:  \begin
{equation}  \label   {g55}   \begin   {array}   {ll}   \hspace   {2mm}
\displaystyle \frac {d(c{\bf p})}{dt}=ce{\bf E}+e{\bf v}{\rm x}{\bf H}
\to & \displaystyle m _ o\frac {d {\bf v}} {dt} = \frac {c} {c _ o}  e
{\bf E} + \frac {e} {c _ o} {\bf v} {\rm x} {\bf H};  \\ \displaystyle
\frac{d {\it E}} {dt} = e {\bf v} \cdot {\bf E} \to & \displaystyle  m
_  o\frac  {dc}  {dt} = \frac {e} {c _ o} {\bf v} \cdot {\bf E}.  \end
{array} \end {equation} From here it can be seen that the integrals of
motion  are either the generalized momentum $ c {\bf p} $ and energy $
E $, or the associated velocity of a particle $ {\bf v}$ and the speed
of light $ c $ in the absence of external forces.

\subsection {Maxwell   Equations  in  LSR}  Taking  into  account  the
expression for the speed of light (\ref {fc}), we obtain the following
form  of Maxwell equations \cite{Kot68}:  \begin {equation} \label {g}
\begin {array} {ll} \vspace {2mm} \displaystyle \nabla {\rm X} {\bf E}
+  \frac  {1}  {c  _  0\sqrt  {1 + \frac {v ^ 2} {{c _ 0} ^ 2}}} \frac
{\partial {\bf H}} {\partial t} = 0; & \nabla\cdot {\bf E} = 4\pi\rho;
\\  \vspace {2mm} \displaystyle \nabla {\rm X} {\bf H} -\frac {1} {c _
0\sqrt {1 + \frac {v ^ 2} {{c _ 0} ^ 2}}}  \frac  {\partial  {\bf  E}}
{\partial  t} = 4\pi\rho\frac {{\bf v}} {c _ 0\sqrt {1 + \frac {v ^ 2}
{{c _ 0} ^ 2}}};  &  \nabla\cdot  {\bf  H}  =  0.  \end  {array}  \end
{equation} Here $ {\bf v} $ is the electrical charge velocity; $ c = c
_ 0\sqrt {1 + v ^ 2 / {c _ 0} ^ 2} $ is the charge coordinate  on  the
axis  $ c $ ($c$ is the speed of light in the laboratory frame $ K $);
$ c _ 0 $ is the proper value of the speed of light in the frame  $  K
$.

\subsection {LSR  and  Experiment}  Let  us  consider  the examples of
experiments, the interpretation of which is close to or coincides with
their interpretation in SR.

{\bf The Michelson Experiment} \cite {Pau47, Lan76}. For the case of a
terrestrial light source the negative result of the experiment may  be
explained  by  space  isotropy  (the speed of light is the same in all
directions).  Owing to this circumstance the interference pattern will
not  be  changed  for  a  terrestrial  observer  at  rotation  of  the
interferometer.  In the case of a extraterrestrial  light  source  the
negative  result  may be explained by two factors:  the space isotropy
and the square dependence of the speed of light from the velocity  $V$
of  a  light source $ c = c _ o\sqrt {1 + V ^ 2 / {c _ o} ^ 2} $ \cite
{Kot68}.

{\bf The Fizeau Experiment} \cite {Pau47,  Lan76}.  The explanation is
similar  to  the one accepted in SR.  The arising little correction is
the the order of $ V ^ 2 / {c _ 0} ^ 2\ll 1 $ and does  not  influence
on  the  experimental  result  in  linear approximation \cite {Kot68}.
(Here $ V $ is the velocity of fluid).

{\bf The Bonch-Bruevich and Molchanov Experiment} \cite  {Bon56}.  The
authors  compared  the speeds of the light radiated by the eastern and
western equatorial edge of the solar disk. In the framework of LSR the
speed  of  light $ c = c _ 0\sqrt {1 + V ^ 2 / {c _ 0} ^ 2} $ does not
depend on the direction of the light source motion $  V  $.  Therefore
the  speed  of light will be the same for both the western and eastern
edges of the solar disk.  As in SR it is in accord with  the  negative
result of the experiment \cite {Kot68}.

\begin {sloppypar}  {\bf  The Sadeh Experiment} \cite {Sad63}.  In the
experiment the distinction between the speeds of the gamma  -  quanta,
arising as a result of the electron - positron annihilation in flight,
has been observed depending on the angle between the gamma  -  quanta.
By  virtue  of  the  independence of the speed of light $ c $ from the
direction of the velocity of the source $ {\bf V} $, the result of the
experiment  should  be negative in LSR as well as in SR \cite {Kot68}.
\end {sloppypar}
\bigskip

\noindent Let  us consider also the experiments,  which interpretation
in LSR is different from their the interpretations in SR. \bigskip

\begin{sloppypar}{\bf The Doppler Effect} \cite {Pau47, Lan60}. In LSR
the  change  of  a  wavelength $\lambda$ is described by the formula $
\lambda = \lambda _ 0 (1-Vn _ x/c) / \sqrt {1-V ^ 2/c ^ 2} = \lambda _
0  [\sqrt  {1 + V ^ 2/ {c _ 0} ^ 2} -Vn _ x/c _ 0] $.  The change of a
frequency is described by the formula $ \omega = \omega _ 0 (c/c _  0)
\sqrt  {1-V ^ 2/c ^ 2} / (1-Vn _ x/c) = \omega _ 0/ ( 1-Vn _ x/ c _ 0)
$ \cite {Kot68}.  Here $ \theta = arc cos n _ x $ is the angle of  the
observation;  $  V $ is the emitter velocity.  It is follows from here
that in LSR there is no Doppler  transversal  frequent  shift  because
with  $  n  _ x = 0 $,  $ \omega = \omega _ o $.  For a wavelength the
Doppler transversal shift is retained.  Hence in LSR the parameters of
redshifts  $ z _ {\lambda} $ and $ z _ {\omega} $ do not coincide with
each other and are equal $ z  _  {\lambda}  =  (\lambda-\lambda  _  0)
/\lambda  _ 0\to 2V / {c _ 0} $ at $ n _ x = -1,  V\to\infty $;  $ z _
{\omega} = (\omega _ 0-\omega) /\omega _ 0\to 1 $ at $ n  _  x  =  -1,
V\to\infty  $.  In non - relativistic approximation they coincide with
each other $ z _ {\lambda} \sim-Vn _ x/c _ o $, $ z _ {\omega} \sim-Vn
_ x/c _ o $.  When the emitters move with significant velocities,  the
distinction begins to show itself with the shifts $ z _ {\lambda} \geq
0.6 $. The fulfillment of the inequality $ z _ {\lambda}\geq\sqrt2 $ =
1.41 is the criterion for longitudinal ($ n _ x = -1  $)  superluminal
motion.  The fulfillment of the inequality $ z _ {\lambda}\geq\sqrt2-1
$ = 0.41 is the criterion for transversal ($ n _ x = 0 $) superluminal
motion \cite {Kot68}.  The superluminal quasars 3C279 ($ z _ {\lambda}
$ = 0.536),  3C345 ($ z _ {\lambda} $ = 0.595), 3C179 ($ z _ {\lambda}
$ = 0.846),  NRAO 140 ( $ z _  {\lambda}  $  =  1.258)  \cite  {Sci81}
satisfy   the   letter   criterion.   In  particular,  the  calculated
transversal velocity of the QSO NRAO 140 expansion is $  V  _  {\perp}
\sim2c _ o $.  It is surprising that this velocity is close to the low
bound of these velocities $ 3c _ o $ within the Friedmann cosmological
model \cite {Sci81}. It is important for LSR to determine the frequent
redshifts $ z _ {\omega} $ of these superluminal objects  and  compare
them  with  the lambda redshifts $ z _ {\lambda} $ as well as to solve
the problem of the existence of the limit $ z _ {\omega} \leq 1 $.  It
will  permit  one  to distinguish between LSR and SR because in SR the
ratio $ z _ {\lambda} = z _ {\omega} $ is true.  The redshift  of  the
radio  emission  from  neutral  hydrogen  $  H  _ I $ on the frequency
corresponding to the line  21  cm  is  attractive  for  this  purpose.
However  in this frequency range the experimental data on superluminal
quasars are not avallable.  Therefore to reject LSR  is  not  possible
now. \end{sloppypar}

Let us also pay attention to the relationship between  the  speeds  of
light  $ c $ and $ c _ o $ and the parameters of redshifts $ c = c _ 0
(1 + z _ {\lambda}) / (1 + z _ {\omega}) \to c _ 0 (1 + z _ {\lambda})
/2 $ in LSR. (The latter formula is true with $ z _ {\omega}\sim 1 $).
We can conclude that it is the  Loiseau  formula  \cite  {Loi68}.  Its
application  to  the  observational  interpretation  was considered in
Introduction.  According to \cite{Loi68},  the speed of light from the
galaxy  NGC  5668  with  the parameters of redshifts $ z _ {\lambda} =
0.00580 $ and $ z _ {\omega} = 0.00526 $,  is equal $ c  =  c  _  0  +
182.04  $  km  /  sec.  In  the light of the present work this result,
however, is not of statistical significance, as the redshift parameter
$ z _ {\lambda} = 0.00580 << 1.41 $. Therefore the conclusion that the
NGC 5668 galaxy is superluminal fail. It also holds for the quasar PKS
2134  with  parameter  of  redshift  $ z _ {\lambda} = 1.935 $.  After
putting this value and frequency shift $ z _ {\omega} = 1 $ into above
-  mentioned  formula we can conclude that the speed of light from the
PKS 2134 quasar is $ c= 300.000\cdot 2.936/2 = 440.400\sim  440.000  $
km / sec. Thus, the Loiseau estimation has theoretical character. This
circumstance indicates once  more  that  it  is  necessary  to  obtain
experimental data concerning the redshifts for superluminal quasars in
the radio-frequency and optical ranges \cite {Kot68}.

{\bf Aberration of light} \cite {Pau47,  Lan76}. By analogy with SR we
have  for  one  -  half  of  the aberration angle:  $ sin\alpha = V/c;
\alpha\sim ( V/c _ 0) (c _ 0/c) = 10 ^ {-4} (c _ 0/c) \sim 10  ^  {-4}
(2  /  (1  +  z  _ {\lambda})) = (2 / (1 + z _ {\lambda})) \cdot20,5 $
seconds of arc.  (The latter is true with large $ z _ {\lambda} $). It
follows from here that, for example, \ $ c = 2,86c _ 0, \ \alpha = 7.2
$ seconds of arc for the Q 1158 + 4635 quasar with the redshift $ z  _
{\lambda} $ = 4.73 \cite {Car90}.  The 7.2 seconds of arc value should
be checked in the experiment \cite {Kot68}.

{\bf Superluminal  motion}  of   nuclear   reaction   products.   Such
phenomenon  is  impossible  in SR.  But it is possible in LSR,  if the
energy of a particle will be greater than $ \sqrt2 E _ 0 $.  It is 150
MeV  for  $  \mu  $  - mesons.  Therefore in LSR (if it is realized in
reality) the appearance of atmospheric $  \mu  $  -  mesons  near  the
surface  of  the  Earth may be explained by superluminal motion of the
mesons with the velocity of the order of $ 6\cdot10 ^  6/2,2\cdot10  ^
{-6} \sim3\cdot10 ^ {12} $ cm/sec,  or $ 100c _ 0 $ \cite {Kot68}. The
energy $ E _ {\mu} = m _ {0,  \mu} c _ 0c\sim100m _ {0, \mu} {c _ 0} ^
2\sim10.6 $ GeV corresponds to the given velocity in LSR. In virtue of
the absence of the limitation on the  upper  value  of  the  speed  of
light,  faster particles explaining the results of Clay,  Crouch \cite
{Cla74} and Cooper experiments \cite {Coo79 *},  may  be  observed  in
front of the particles from nuclear reactions.

\begin {sloppypar}   {\bf   Motion   of   a   charged   particle}   in
electromagnetic field.  By integrating (\ref {2.33}),  we find that in
the  case  a  particle moves in constant homogeneous electrical field,
its velocity tends to infinity $ v _ x (t) = c _ 0\sqrt {1 + {v _ y} ^
2  (0)  /  {c  _ 0} ^ 2} sh (e {\rm E} t/m _ 0c _ 0) \to\infty $ \cite
{Kot68}.  In SR the particle velocity is limited by value $ c _ o $ as
is  known \cite{Lan60}.  For the case of constant homogeneous magnetic
field $ {\bf H} = (0,0,{\rm H _ z}) $ the frequency of rotation  of  a
particle is constant and does not depend on the energy of a particle $
\omega = e {\rm H _ z} / {m _ 0c _ 0} $ = const in LSR  ($\omega  \sim
1/E$ in SR \cite{Lan60}). However if the particle energy is great, the
radius of particle rotation is connected with the particle  energy  by
the  ratio $ r\sim {\it E} /eH _ z $ as in SR.  The differences in the
radiuses of rotation is observed in the intermediate range of particle
energies,  when \ $ M _ 0 {c _ 0} ^ 2 < {\it E} < m _ 0c _ 0v $ \ at $
v >> c _ o  $.  The  considered  properties  of  particle  motions  in
electrical  and  magnetic  fields  may  be  essential in the theory of
linear and cyclical accelerators \cite {Kot68}. \end {sloppypar}

\section{Conclusion} Summing we shall note that the validity of LSR or
the proper field of its application are not clear yet now. In any case
the problem  arises  which  concerns  the  reason  of  the  choice  of
preferable symmetry in the nature. Local SR transforms into SR, if $ c
' = c $.

 \end {document}